\documentclass[twocolumn,english,showkeywords]{revtex4}
\usepackage[T1]{fontenc}
\usepackage[latin1]{inputenc}
\usepackage{graphicx}
\usepackage{amssymb}

\makeatletter

\renewcommand{\marginpar}[1]{}

\newcommand{\etal}[1]{, #1}

\usepackage{babel}
\makeatother
\begin{document}
\newcommand{\mubohr}{{\mu_{\mathrm{B}}}}

\newcommand{\efermi}{{\varepsilon_{\mathrm{F}}}}

\newcommand{\bra}[1]{\langle#1|}

\newcommand{\ket}[1]{|#1\rangle}

\newcommand{\ibra}[1]{{}_{\mathrm{I}}\bra{#1}}

\newcommand{\iket}[1]{\ket{#1}_{\mathrm{I}}}

\newcommand{\rotLbl}{{\mathrm{r}}}

\newcommand{\rotbra}[1]{{}_{\rotLbl}\bra{#1}}

\newcommand{\rotket}[1]{\ket{#1}_{\rotLbl}}

\newcommand{\braket}[2]{{\langle#1|#2\rangle}}

\newcommand{\ketbraUnity}[1]{\ket{#1}\bra{#1}}

\newcommand{\braTwo}[2]{\bra{#1,\, \hat{#2}}}

\newcommand{\ketTwo}[2]{\ket{#1,\, \hat{#2}}}

\newcommand{\expect}[1]{\left\langle #1\right\rangle }

\newcommand{\spS}{\ket{S}}

\newcommand{\spup}{\ket{\! \uparrow}}

\newcommand{\spdown}{\ket{\! \downarrow}}

\newcommand{\spupbra}{\bra{\uparrow\! }}

\newcommand{\spdownbra}{\bra{\downarrow\! }}

\newcommand{\spupup}{\ket{\! \uparrow\uparrow}}

\newcommand{\spupdown}{\ket{\! \uparrow\downarrow}}

\newcommand{\spdownup}{\ket{\! \downarrow\uparrow}}

\newcommand{\spdowndown}{\ket{\! \downarrow\downarrow}}

\newcommand{\du}{{\downarrow\uparrow}}

\newcommand{\ud}{{\uparrow\downarrow}}

\newcommand{\rhoElem}[1]{\rho_{#1}}

\newcommand{\rhoDotElem}[1]{{\dot{\rho}_{#1}}}

\newcommand{\rhoDot}{{\rhoDotElem{}}}

\newcommand{\rhoDotuu}{{\rhoDotElem{\uparrow}}}

\newcommand{\rhoDotdd}{{\rhoDotElem{\downarrow}}}

\newcommand{\rhoDotdu}{{\rhoDotElem{\downarrow,\uparrow}}}

\newcommand{\rhouu}{{\rhoElem{\uparrow}}}

\newcommand{\rhodd}{{\rhoElem{\downarrow}}}

\newcommand{\rhoSS}{{\rhoElem{S}}}

\newcommand{\rhoud}{\rhoElem{\uparrow,\downarrow}}

\newcommand{\rhodu}{\rhoElem{\downarrow,\uparrow}}


\newcommand{\gammaLS}[1]{\gamma_{ls,\, {#1}}}

\newcommand{\sqrtGammaLS}[1]{\Gamma_{ls,\, {#1}}}

\newcommand{\gammaTZero}{\gamma}

\newcommand{\gal}[1]{\gammaTZero_{#1}^{\alpha}}

\newcommand{\gmal}[1]{\gammaTZero_{#1}^{-\alpha}}

\newcommand{\gl}[1]{\gammaTZero_{#1}}

\newcommand{\gul}[1]{\gammaTZero_{#1}^{\uparrow}}

\newcommand{\gdl}[1]{\gammaTZero_{#1}^{\downarrow}}

\newcommand{\gull}{\gul{l}}

\newcommand{\gdll}{\gdl{l}}

\newcommand{\guO}{\gul{1}}

\newcommand{\guT}{\gul{2}}

\newcommand{\gdO}{\gdl{1}}

\newcommand{\gdT}{\gdl{2}}

\newcommand{\gO}{\gammaTZero_{1}}

\newcommand{\gT}{\gammaTZero_{2}}

\newcommand{\Dm}{\Delta\mu}

\newcommand{\HDD}{H_{\mathrm{DD}}}

\newcommand{\HDL}{H_{\mathrm{DL}}}

\newcommand{\tDD}{t_{\mathrm{DD}}}

\newcommand{\tDL}{t_{\mathrm{DL}}}

\newcommand{\sysLbl}{}

\newcommand{\rhoSys}{\rho_{\sysLbl}}

\newcommand{\rhoDotSys}{\dot{\rho}_{\sysLbl}}

\newcommand{\rhoStat}{\bar{\rho}}

\newcommand{\rhoFull}{\rho_{\mathrm{F}}}


\newcommand{\TrB}{{{\mathrm{Tr}}_{\bathLbl}\, }}

\newcommand{\TrS}{{{\mathrm{Tr}}_{\sysLbl}\, }}

\newcommand{\TrF}{{{\mathrm{Tr}}\, }}

\newcommand{\LS}{{L_{\sysLbl}}}

\newcommand{\LBath}{{L_{\bathLbl}}}

\newcommand{\HBath}{{H_{\bathLbl}}}

\newcommand{\LNot}{L_{0}}

\newcommand{\LV}{{L_{V}}}


\newcommand{\decoherenceSymbol}{V}

\newcommand{\rateSymbol}{W}

\newcommand{\effectiveRate}[1]{\rateSymbol_{#1}}

\newcommand{\effectiveRateMax}[1]{W_{#1}^{\mathrm{max}}}

\newcommand{\Xu}{X\uparrow}

\newcommand{\Xd}{X\downarrow}

\newcommand{\XdXu}{\Xd,\Xu}

\newcommand{\XuXd}{\Xu,\Xd}

\newcommand{\XuXu}{\Xu}

\newcommand{\XdXd}{\Xd}

\newcommand{\dXd}{\downarrow,\Xd}

\newcommand{\uXd}{\uparrow,\Xd}

\newcommand{\dXu}{\downarrow,\Xu}

\newcommand{\uXu}{\uparrow,\Xu}

\newcommand{\Xdd}{\Xd,\downarrow}

\newcommand{\Xdu}{\Xd,\uparrow}

\newcommand{\Xud}{\Xu,\downarrow}

\newcommand{\Xuu}{\Xu,\uparrow}

\newcommand{\exLabel}{\mathrm{L}}

\newcommand{\exNot}{X^{0}}

\newcommand{\exM}{X^{-}}

\newcommand{\exMket}{\ket{\exM}}

\newcommand{\exMbra}{\bra{\exM}}

\newcommand{\exMup}{X_{\uparrow}^{-}}

\newcommand{\exMupket}{\ket{\exMup}}
 
\newcommand{\exMupbra}{\bra{X_{\uparrow}^{-}}}

\newcommand{\exMdown}{X_{\downarrow}^{-}}

\newcommand{\exMdownket}{\ket{\exMdown}}
 
\newcommand{\exMdownbra}{\bra{\exMdown}}

\newcommand{\RabiEx}{\Omega_{\exLabel}}

\newcommand{\detuningEx}{\delta_{\exLabel}}

\newcommand{\WeffEx}{\effectiveRate{\exLabel}}

\newcommand{\decohEx}{\decoherenceSymbol_{\mathrm{X}}}

\newcommand{\omegaESR}{\omega_{\mathrm{ESR}}}

\newcommand{\RabiESR}{\Omega_{\mathrm{ESR}}}

\newcommand{\detuningESR}{\delta_{\mathrm{ESR}}}

\newcommand{\WeffESR}{\effectiveRate{\mathrm{ESR}}}

\newcommand{\WmaxESR}{\effectiveRateMax{\mathrm{ESR}}}

\newcommand{\decohESR}{\frac{1}{T_{2}}}

\newcommand{\detuningESReff}{\tilde{\delta}_{\mathrm{ESR}}}

\newcommand{\decohESReff}{\decoherenceSymbol_{\mathrm{ESR}}}

\newcommand{\Wud}{W_{\ud}}

\newcommand{\Wdu}{W_{\du}}

\newcommand{\WSud}{\tilde{W}_{\ud}}

\newcommand{\WSdu}{\tilde{W}_{\du}}

\newcommand{\Wpceff}{\tilde{\rateSymbol}_{\mathrm{pc}}}

\newcommand{\WXdXu}{\rateSymbol_{\XdXu}}

\newcommand{\WXuXd}{\rateSymbol_{\XuXd}}

\newcommand{\Wem}{\rateSymbol_{\mathrm{em}}}

\newcommand{\Wpc}{\rateSymbol_{\mathrm{pc}}}

\newcommand{\Wswitchoff}{\rateSymbol_{\mathrm{r}}}

\newcommand{\bathLbl}{\mathrm{R}}

\newcommand{\HSrot}{H'}

\newcommand{\Hdot}{H_{\mathrm{dot}}}

\newcommand{\HphotonField}{H_{\mathrm{L}}}

\newcommand{\photolum}[1]{\Gamma^{#1}}

\newcommand{\reptime}{\tau_{\mathrm{rep}}}

\newcommand{\rhoinf}{\rho_{\infty}}

\newcommand{\meq}{\mathcal{M}}

\newcommand{\meqL}{\meq_{\mathrm{L}}}

\newcommand{\meqnull}{\meq_{0}}

\title{Probing Single-Electron Spin Decoherence in Quantum Dots using Charged
Excitons}

\author{Oliver Gywat}

\thanks{Department of Physics and Astronomy, University of Basel, Klingelbergstrasse
82, CH-4056 Basel, Switzerland}

\author{Hans-Andreas Engel}

\thanks{Department of Physics and Astronomy, University of Basel, Klingelbergstrasse
82, CH-4056 Basel, Switzerland}

\author{Daniel Loss}

\thanks{Department of Physics and Astronomy, University of Basel, Klingelbergstrasse
82, CH-4056 Basel, Switzerland}

\begin{abstract}
We propose to use optical detection of magnetic resonance (ODMR) to
measure the decoherence time $T_{2}$ of a single electron spin in
a semiconductor quantum dot. The electron is in one of the spin 1/2
states and a circularly polarized laser can only create an optical
excitation for one of the electron spin states due to Pauli blocking.
 An applied electron spin resonance (ESR) field leads to Rabi spin
flips and thus to a modulation of the photoluminescence or, alternatively,
of the photocurrent. This allows one to measure the ESR linewidth
and the coherent Rabi oscillations, from which the electron spin decoherence
can be determined. We study different possible schemes for  such
an ODMR setup, including cw or pulsed laser excitation.  
\end{abstract}

\maketitle

\section{Introduction}

The spin $1/2$ states $\spup$ and $\spdown$ of an electron in a
semiconductor quantum dot can be used as an implementation of a quantum
bit (qubit) \cite{spintronics}. Due to the rather weak coupling of
the spin and orbital degrees of freedom in quantum dots, the electron
spin is only weakly interacting with its environment. Still, there
is a finite lifetime for spin in these systems, limiting the time
during which quantum information can be processed. For single spins,
one distinguishes between the two characteristic decay times $T_{1}$
and $T_{2}$. The relaxation of an excited spin state in a magnetic
field into the thermal equilibrium occurs with the spin relaxation
time $T_{1}$, whereas the spin decoherence time $T_{2}$ is associated
with the loss of phase coherence of a single spin that is prepared
in a coherent superposition of its eigenstates. Recent experiments
with InGaAs and GaAs dots have shown extremely long spin relaxation
times up to $T_{1}\approx1\:\mathrm{ms}$ \cite{fujisawa,hanson,kouwenhoven}.
Experimental $T_{2}$ measurements of single electron spins in quantum
dots are highly desirable and have not been accomplished so far. It
would also be interesting to experimentally verify the theoretical
prediction $T_{2}=2T_{1}$ for a quantum dot (with decoherence due
to the spin-orbit interaction) \cite{golovach}.

Coherent control and detection of excitonic states in single quantum
dots has been demonstrated in optical experiments \cite{gammon1}.
Nevertheless, the $T_{2}$ time of a single electron spin in a quantum
dot has not yet been measured successfully using optical methods.
In this respect, the interaction of the electron and the hole of the
exciton imposes a principal difficulty:    The electron spin and
the hole spin are only decoupled if the hole spin couples stronger
to the environment than to the electron spin.   However, time-resolved
Faraday rotation experiments suggest that there is significant coupling
of electron and hole spins in quantum dots \cite{guptaT2}.   In
many other experiments, electron-hole pairs are excited inside the
barrier material of a quantum dot heterostructure. After their creation,
the carriers diffuse into the dots within typically tens of picoseconds
\cite{ohnesorge,raymond}. Due to the fast relaxation time of the
hole spin in the barrier, electron and hole spins decouple during
this time.  One would thus expect that in such a setup only the spin
decoherence of electrons can be measured, e.g., by the  Hanle effect
\cite{epstein}. But this approach has not yet given conclusive results
for $T_{2}$. Alternatively, $T_{2}$ can be measured via currents
through quantum dots in an ESR field \cite{engel:01,engel:02,martin}.
However, this requires contacting of the dots with current leads which
reduces coherence, while with an optical detection scheme one can
also benefit from the high sensitivity of photodetectors. 

Optical detection of magnetic resonance (ODMR) has already been applied
to measure the coherence of single spins in various systems, including
single molecules \cite{koehler,wrachtrup} and single nitrogen-vacancy
centers in diamond \cite{gruber,jelezko}. Recent ODMR experiments
on charge-neutral semiconductor quantum dots \cite{Lifshitz,zurauskiene}
have demonstrated the feasibility of the combination of ESR and optical
methods in quantum dot experiments, but have not considered single
spin coherence.   

In this article, we start by reviewing our recent proposal \cite{gywat}
to measure $T_{2}$ of a single electron spin in a semiconductor quantum
dot via ODMR.  We also add new results on the detection using photocurrent
and on the luminescence intensity autocorrelation function. In Section
\ref{sec:Negatively-Charged-Excitons}, we discuss the states of a
negatively charged exciton in a quantum dot. We introduce the Hamiltonian
in Section \ref{sec:Hamiltonian} and describe the dynamics of the
ODMR setup with a generalized master equation in Section \ref{sec:Generalized-Master-Equation}.
We show in Section \ref{sec:ESR-Linewidth-in} that the linewidth
of the photoluminescence as function of the ESR frequency provides
a lower bound on $T_{2}$ \cite{gywat}. Extending our previous work,
we elaborate on the read out via photocurrent in Section \ref{sec:Read-Out-Via}
and discuss in Section \ref{sec:Luminescence-Intensity-Autocorrelation}
the autocorrelation function of the luminescence intensity as another
possible detection scheme. Further, we identify a regime where $T_{1}$
can be measured optically. We show in Section \ref{sec:Spin-Rabi-Oscillations}
that electron spin Rabi oscillations can be detected via the photoluminescence
if pulsed laser and cw ESR excitation is applied. Using pulsed laser
excitation, electron spin precession can be detected with similar
schemes, as we discuss in Section \ref{sec:Spin-precession-via}.
We conclude in Section \ref{sec:Conclusions}.

\section{Negatively Charged Excitons\label{sec:Negatively-Charged-Excitons}}

\begin{figure}[tb]
 \centerline{\includegraphics[%
  width=8.6cm,
  keepaspectratio]{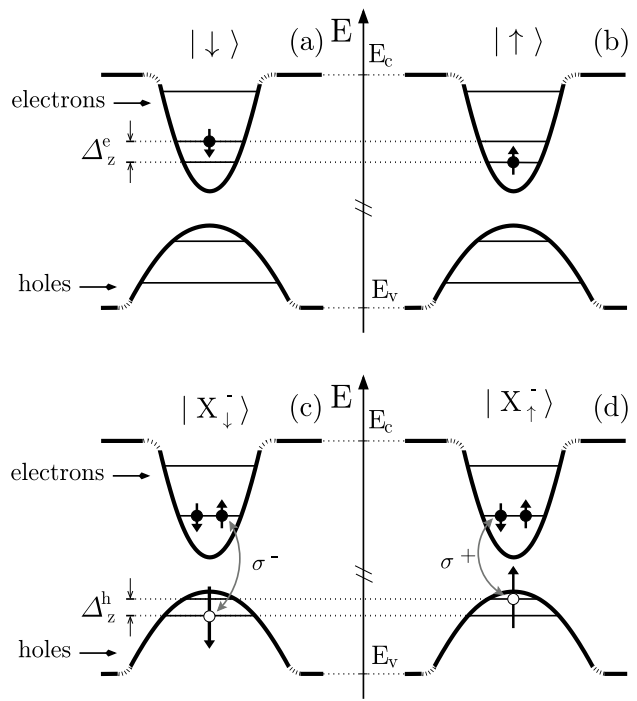} \vspace{1mm} }

\caption{(a)-(d): The states of a single quantum dot in a static magnetic
field. The Zeeman splittings are $\Delta_{z}^{e}=g_{e}^{z}\mubohr B_{z}$
for the electron and $\Delta_{z}^{h}=g_{hh}^{z}\mubohr B_{z}$ for
the hole.  Coherent transitions occur between (a) and (b) due to
an ESR field and between (a) and (c) due to a $\sigma^{-}$-polarized
laser field.  The grey arrows in (c) and (d) indicate which electron-hole
pair couples with the photon field of polarization $\sigma^{\pm}$.
\label{fig:states}}
\end{figure}

We consider a quantum dot that confines electrons as \emph{}well as
\emph{}holes (i.e., a type I dot). We assume that the dot is charged
with one single electron. This can be achieved experimentally, e.g.,
by n-doping \cite{cortez}, or by electrical injection if the dot
is embedded inside a photodiode structure \cite{abstreiter2,schulhauser}.
The single-electron state of the dot can be optically excited, creating
a negatively charged exciton ($X^{-}$) which consists of two electrons
and one hole. In the $X^{-}$ ground state, the two electrons form
a spin singlet in the lowest (conduction-band) electron level and
the hole occupies the lowest (valence-band) hole level, as shown in
recent experiments with InAs dots \cite{trion1,trion2} and GaAs dots
\cite{trion3}.  Such negatively charged excitons can be used to
read out and initialize a single electron spin \cite{shabaev}. 
We assume that the lowest heavy hole (hh) dot level (with total angular
momentum projection $J_{z}\!=\!\pm3/2$) and the lowest light hole
(lh) dot level (with $J_{z}\!=\!\pm1/2$) are split by an energy $\delta_{hh-lh}$
and that  mixing of hh and lh states is negligible. These conditions
are satisfied for several types of quantum dots \cite{trion1,trion2,trion3,shabaev,efrosCdSe,efros2}.
Then, if excitation is restricted to either hh or lh states, the circularly
polarized optical transitions $\sigma^{+}$ and $\sigma^{-}$ are
unambigously related to one spin polarization of the conduction-band
electron because of optical selection rules, see also Fig.\  \ref{fig:states}.
Here, we assume a hh ground state for holes.       %

For the proposed ODMR scheme, we consider the following dot states
(see also Fig.\  \ref{fig:states}). In the presence of an external
static magnetic field, a single electron in the lowest orbital state
can be in the spin ground state $\spup$ or in the excited spin state
$\spdown$. Similarly, an $X^{-}$ in the orbital ground state can
either be in the excited spin state $\exMdownket$ or in the spin
ground state $\exMupket$, where the subscripts $\downarrow$ and
$\uparrow$ refer to the hh spin. We apply \textbf{}the usual time-inverted
\textbf{}notation for hole spins. For simplicity, we have assumed
equal signs for the electron and the hh $g$ factors in $z$ direction.
   Here, we exclude $X^{-}$ states where one electron is in an
excited orbital state. The lowest $X^{-}$ state of this type contains
an electron triplet and requires an additional energy $\delta\epsilon\approx40$
meV in InAs dots \cite{cortez}. This energy difference $\delta\epsilon$
is mainly given by the single-electron level spacing ($\approx50$
meV \cite{fricke}) and the electron-electron exchange interaction.
 Consequently, the state $\exMdownket$ can be excited resonantly
by a circularly polarized laser with a bandwidth lower than $\delta\epsilon$
and $\delta_{hh-lh}$.   An ODMR scheme including an $X^{-}$ state
with an excited hole is also possible, as we discuss in Section \ref{sec:ESR-Linewidth-in}.

\section{Hamiltonian\label{sec:Hamiltonian}}

       In an ODMR setup with a quantum dot containing a single
excess electron, we describe the energy-conserving dynamics with the
Hamiltonian \begin{equation}
H=\Hdot+H_{\mathrm{ESR}}+\HphotonField+H_{\mathrm{d-L}},\label{eq:H}\end{equation}
 which couples the three states $\spup$, $\spdown$, and $\exMdownket$.
 Here, $\Hdot$ contains the quantum dot potential, the Zeeman energies
due to a constant magnetic field $B_{z}$ in the $z$ direction, and
the Coulomb interaction of electrons and holes. The dot energy $E_{n}$
is defined by $\Hdot\ket{n}=E_{n}\ket{n}$. We set $\hbar=1$ in the
following. The electron Zeeman splitting is $\Delta_{z}^{e}=g_{e}^{z}\mubohr B_{z}=E_{\downarrow}-E_{\uparrow}$,
where $g_{e}^{z}$ is the electron $g$ factor and $\mubohr$ is the
Bohr magneton. In $B_{z}$, we also include the Overhauser field 
which could possibly arise from dynamically polarized nuclear spins.
 The ESR term $H_{\mathrm{ESR}}(t)$ couples the two electron Zeeman
levels $\spup$ and $\spdown$ via the magnetic field $\mathbf{B}_{\perp}(t)$,
which rotates with frequency $\omegaESR$ in the $xy$ plane. Note
that a linearly oscillating magnetic field, $\mathbf{B}_{x}(t)=B_{x}^{0}\cos(\omegaESR t)\mathbf{\hat{x}}$,
can be applied instead of $\mathbf{B}_{\perp}(t)$ \cite{sakurai}.
In the rotating wave approximation, this field leads to the same result
as the rotating field with $B_{\perp}\rightarrow B_{x}^{0}/2$. The
ESR Rabi frequency is $\RabiESR=g_{e}^{\perp}\mubohr B_{\perp}$,
with in-plane $g$ factor $g_{e}^{\perp}$ (typically, $g_{e}^{\perp}=g_{e}^{z}$).
Even if the ESR field is also resonant with the hole Zeeman splitting,
the Rabi oscillations of the holes have a negligible effect since
the charged exciton states recombine quickly. As an alternative to
an ESR field, an oscillating field $\mubohr\tensor{\mathbf{g}}\mathbf{B}$
could also be produced using voltage-controlled modulation of the
electron $g$\emph{-}tensor $\tensor{\mathbf{g}}$, which has already
been achieved experimentally in quantum wells \cite{kato}. A $\sigma^{-}$-polarized
laser of frequency $\omega_{\exLabel}$ is applied in $z$ direction
(typically parallel to $[001]$), with the free laser field Hamiltonian
$\HphotonField=\omega_{\exLabel}a_{\exLabel}^{\dagger}a_{\exLabel}$,
where $a_{\exLabel}^{(\dagger)}$are photon operators. The optical
interaction term $H_{\mathrm{d}-\mathrm{L}}$ describes the coupling
of $\spdown$ and $\exMdownket$ to the laser field  with the complex
optical Rabi frequency $\RabiEx$ \cite{gywat}. We take the coupling
to the laser into account in $H_{\mathrm{d-L}}$. Because the laser
is circularly polarized, the terms that violate energy conservation
vanish due to selection rules.   Further, the absorption of a $\sigma^{-}$
photon in the spin ground state $\spup$ is excluded due to Pauli
blocking because we assume that the laser bandwidth is smaller than
$\delta_{hh-lh}$ and $\delta\epsilon$, as discussed in Section \ref{sec:Negatively-Charged-Excitons}.
Note that the very same scheme can also be applied if the sign of
the hole $g$ factor is reversed, since a $\sigma^{+}$ laser field
can then be used and all results apply after interchanging $\exMdownket$
and $\exMupket$. The laser bandwidth and also the temperature can
safely exceed the electron Zeeman splitting.   Finally, we exclude
all multi-photon processes via other levels since they are only relevant
to high-intensity laser fields. In this configuration, the $\sigma^{-}$
photon absorption is switched {}``on'' and {}``off'' by the electron
spin flips driven by the ESR. We  next transform $H$ into the rotating
frame with respect to the field frequencies $\omegaESR$ and $\omega_{\mathrm{L}}$.
 We introduce the laser detuning $\detuningEx=(E_{\Xd}-E_{\downarrow})-\omega_{\mathrm{L}}$
and the ESR detuning $\detuningESR=g_{e}^{z}\mubohr B_{z}-\omegaESR$.

\section{Generalized Master Equation\label{sec:Generalized-Master-Equation}}

For the dot dynamics including relaxation and decoherence processes,
we consider the reduced density matrix for the dot, $\rhoSys=\TrB\rhoFull$.
Here, $\rhoFull$ is the full density matrix of the dot and its environment
(or reservoir), i.e., the unobserved degrees of freedom, and $\TrB$
is the trace taken over the reservoir. We take the interaction of
the dot states with the ESR and the laser fields exactly into account
using the Hamiltonian {[}Eq.\ (\ref{eq:H}){]} in the rotating frame.
With a generalized master equation in the Lindblad form, we take the
coupling with the environment (radiation field, nuclear spins, phonons,
spin-orbit interaction, etc.) into account with phenomenological rates.
We use the rates $W_{nm}\equiv W_{n\leftarrow m}$ for the incoherent
transitions from state $\ket{m}$ to $\ket{n}$ and the rates $\decoherenceSymbol_{nm}$
for the decay of the corresponding off-diagonal matrix elements of
$\rhoSys$. These decoherence rates $\decoherenceSymbol_{nm}$ have
the structure $\decoherenceSymbol_{nm}=\frac{1}{2}\sum_{k}\left(W_{kn}+W_{km}\right)+V_{n}+V_{m},$
where the rate $V_{n}+V_{m}$ is usually called the pure decoherence
rate. Further, the electron spin relaxation time is $T_{1}=\left(\rateSymbol_{\ud}+\rateSymbol_{\du}\right)^{-1}$,
with spin flip rates $\rateSymbol_{\ud},\rateSymbol_{\du}$. (In Section
\ref{sec:ESR-Linewidth-in} below, we point out a method to measure
$T_{1}$ in a similar setup as discussed here.)  In the absence of
the ESR field and the laser field, the off-diagonal matrix elements
of the electron spin decay with the (intrinsic) single-spin decoherence
rate $\decoherenceSymbol_{\du}=\frac{1}{2}(\rateSymbol_{\ud}+\rateSymbol_{\du})+\decoherenceSymbol_{\uparrow}+\decoherenceSymbol_{\downarrow}=1/T_{2}$.
Further, the linewidth of the optical $\sigma^{-}$ transition is
denoted by $\decohEx=\decoherenceSymbol_{\Xdd}$.   We use the notation
$\rho_{n}=\bra{n}\rho\ket{n}$ and $\rho_{n,m}=\bra{n}\rho\ket{m}$
for the matrix elements of $\rho$.  In the rotated basis $\spup$,
$\spdown$, $\exMupket$, $\exMdownket$, the generalized master equation
is given by $\rhoDotSys=\mathcal{M}\rhoSys$, where $\mathcal{M}$
is a superoperator. Explicitly, \begin{eqnarray}
\rhoDotuu & = & \RabiESR{\textrm{Im}}\rhodu\!+\!\Wem\rho_{\XuXu}\!+\!\rateSymbol_{\ud}\rhodd\!-\!\rateSymbol_{\du}\rhouu,\label{eq:fullmastereq}\\
\rhoDotdd & = & -\RabiESR{\textrm{Im}}\rhodu+{\textrm{Im}}(\RabiEx^{*}\rho_{\Xdd})+\Wem\,\rho_{\XdXd}\nonumber \\
 &  & +\rateSymbol_{\du}\,\rhouu-\rateSymbol_{\ud}\,\rhodd,\\
\dot{\rho}_{\XdXd} & = & -{\textrm{Im}}(\RabiEx^{*}\rho_{\Xdd})+\rateSymbol_{\XdXu}\,\rho_{\XuXu}\nonumber \\
 &  & -\left(\Wem+\rateSymbol_{\XuXd}\right)\rho_{\XdXd},\\
\dot{\rho}_{\XuXu} & = & \rateSymbol_{\XuXd}\,\rho_{\XdXd}-\left(\Wem+\rateSymbol_{\XdXu}\right)\rho_{\XuXu},\\
\rhoDotdu & = & \frac{i}{2}\RabiESR\left(\rhodd-\rhouu\right)-\frac{i}{2}\RabiEx^{*}\rho_{\Xdu}\nonumber \\
 &  & -\left(i\detuningESR+T_{2}^{-1}\right)\,\rhodu,\\
\dot{\rho}_{\Xdu} & = & \frac{i}{2}\RabiESR\,\rho_{\Xdd}-\frac{i}{2}\RabiEx\rhodu\nonumber \\
 &  & -[i(\detuningESR+\detuningEx)+\decoherenceSymbol_{\Xdu}]\,\rho_{\Xdu},\\
\dot{\rho}_{\Xdd} & = & \frac{i}{2}\RabiESR\rho_{\Xdu}-\frac{i}{2}\RabiEx(\rhodd-\rho_{\XdXd})\nonumber \\
 &  & -(i\detuningEx+\decohEx)\rho_{\Xdd}.\label{eq:fullmastereqfin}\end{eqnarray}

 The remaining (off-diagonal) matrix elements of $\rhoSys$ are decoupled
from these equations and are not further important here. %
\begin{figure}[t]
 \centerline{\includegraphics[%
  width=8.6cm,
  keepaspectratio]{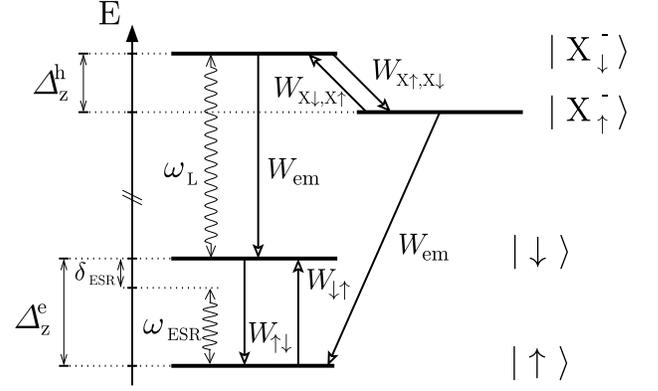} \vspace{1mm} }

\caption{Level scheme of the four states shown in Fig.\ \ref{fig:states}.
Wavy arrows describe the transitions driven by the ESR field and the
laser field with frequencies $\omegaESR$ and $\omega_{\exLabel}$,
respectively. The corresponding Rabi frequencies are $\RabiESR$ and
$|\RabiEx|$. A detuning $\detuningESR=\Delta_{z}^{e}-\omegaESR$
is shown for the ESR field, with Zeeman splitting $\Delta_{z}^{e}$.
Incoherent transitions are shown with  arrows and occur at rates
$\rateSymbol_{nm}$. We consider $\rateSymbol_{\dXd}=\rateSymbol_{\uXu}=:\rateSymbol_{\mathrm{em}}$.
\label{fig:transitions}}
\end{figure}

\section{ESR Linewidth in the Photoluminescence\label{sec:ESR-Linewidth-in}}

We now calculate the stationary photoluminescence for a cw ESR field
and a cw laser field. For this, we evaluate the stationary density
matrix $\rhoStat_{\sysLbl}$, which satisfies $\dot{\rhoStat}_{\sysLbl}=0$.
We introduce the effective rate \begin{equation}
\WeffEx=\frac{|\RabiEx|^{2}}{2}\:\frac{\decohEx}{\decohEx^{2}+\detuningEx^{2}}\label{eqWeffEx}\end{equation}
 for the optical excitation, which takes its maximum value $\WeffEx^{\mathrm{max}}$
for $\detuningEx=0$. We first solve $\dot{\rhoStat}_{\Xdu}=0$. We
find that the coupling to the laser field produces an additional decoherence
channel to the electron spin. We thus obtain a renormalized spin decoherence
rate $\decohESReff$, which satisfies\begin{equation}
\decohESReff\leq\decohESR+\frac{|\RabiEx|^{2}}{4\decoherenceSymbol_{\Xdu}}\approx\decohESR+\frac{1}{2}\WeffEx^{\mathrm{max}}.\label{eqndecohESReff}\end{equation}
 Similarly, the ESR detuning $\detuningESR$ is also renormalized,
\begin{equation}
\detuningESReff\geq\detuningESR\left[1-\frac{|\RabiEx|^{2}}{\left(\Wem+\WXuXd\right)^{2}}\right].\end{equation}
 We assume that these renormalizations and $\detuningEx$ are small
compared to the optical linewidth $\decohEx$, i.e., $\WeffEx^{\mathrm{max}},\,|\detuningESReff-\detuningESR|<\decohEx$.
Further, if both transitions are near resonance, $\detuningEx\lesssim\decohEx$
and $|\detuningESReff|\lesssim\decohESReff$, no additional terms
appear in the renormalized master equation. We then solve $\dot{\rhoStat}_{\Xdd}=0$
and $\dot{\rhoStat}_{\downarrow,\uparrow}=0$ and introduce the effective
Rabi spin-flip rate\begin{equation}
\WeffESR=\frac{\RabiESR^{2}}{2}\:\frac{\decohESReff}{\decohESReff^{2}+{\detuningESReff}^{2}},\label{eqWeffESR}\end{equation}
which together with $\WeffEx$ eliminates the parameters $\RabiEx$,
$\decohEx$, $\detuningEx$, $\RabiESR$, $\decohESReff$, and $\detuningESReff$
in the remaining equations for the diagonal elements of $\rho$. Further,
these now contain the total spin flip rates $\WSud=\Wud+\WeffESR$
and $\WSdu=\Wdu+\WeffESR$. We obtain the stationary solution\begin{eqnarray}
\rhoStat_{\uparrow} & = & \eta\WeffEx\,\Wem\,\WXuXd+\eta\WSud\,\Wem\,\WXuXd\nonumber \\
 &  & +\eta\WSud\left(\WeffEx+\Wem\right)\left(\Wem+\WXdXu\right),\\
\rhoStat_{\downarrow} & = & \eta\WSdu\,\left(\WeffEx+\Wem\right)\left(\Wem+\WXdXu\right)\nonumber \\
 &  & +\eta\WSdu\,\Wem\,\WXuXd,\\
\rhoStat_{\Xd} & = & \eta\WeffEx\,\WSdu\,\left(\Wem+\WXdXu\right),\\
\rhoStat_{\Xu} & = & \eta\WeffEx\,\WSdu\,\WXuXd,\end{eqnarray}
where the normalization factor $\eta$ is chosen such that $\sum_{n}\rho_{n}=1$.
Comparing the expressions for $\rhoStat_{\uparrow}$ and $\rhoStat_{\downarrow}$
above, we see that $\rhoStat_{\uparrow}\geq\rhoStat_{\downarrow}$
is satisfied for $\Wud\geq\Wdu$. This electron spin polarization
is due to the hole spin relaxation channel, analogous as in an optical
pumping scheme. A hole spin flip corresponds to leakage out of the
states that are driven by the external fields. Since the dynamics
due to the ESR is much slower than the optical recombination, there
is an increased population of the state $\spup$.  %
\begin{figure}[t]
\includegraphics[%
  width=8.6cm,
  keepaspectratio]{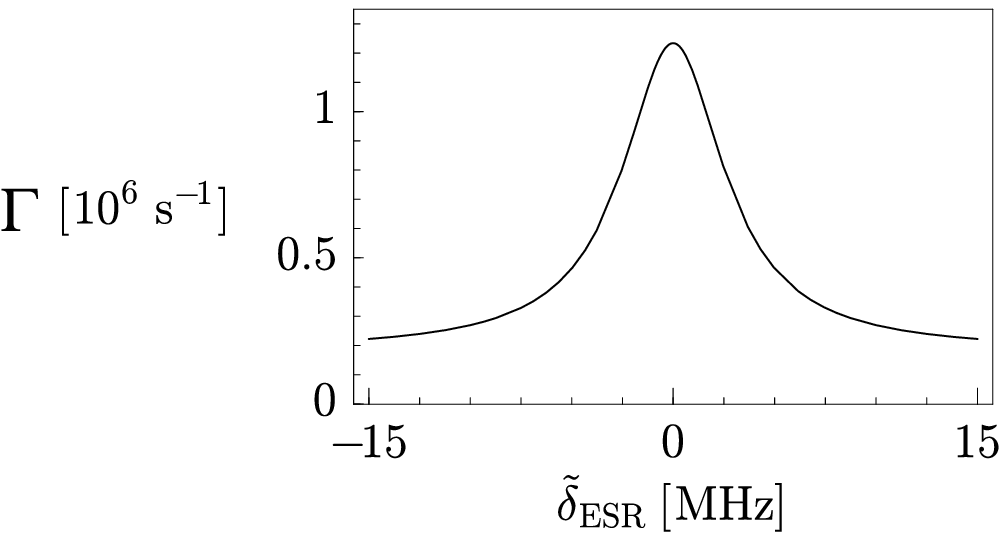}

\caption{\label{Fig: cwPL}The stationary photoluminescence rate $\Gamma$
as a function of the ESR detuning $\detuningESReff$. As described
in the text, $\Gamma$ is a Lorentzian and its linewidth $w$ gives
an upper bound for $2/T_{2}$. Here, we use $g_{e}=0.5$, $B_{\perp}=1\:\mathrm{G}$,
$T_{2}=100\:\mathrm{ns}$, $\rateSymbol_{\ud}=\rateSymbol_{\du}=(20\:\mathrm{\mu s})^{-1}$,
$\rateSymbol_{\mathrm{em}}=10^{9}\:\mathrm{s^{-1}}$, $\rateSymbol_{\XuXd}=\rateSymbol_{\XdXu}=\rateSymbol_{\mathrm{em}}/2$,
$\detuningEx=0$, $\decoherenceSymbol_{\Xdu}=\decohEx=(\rateSymbol_{\mathrm{em}}+\rateSymbol_{\XuXd})/2$,
and $\RabiEx=2\RabiESR\sqrt{T_{2}\decohEx}$. With these parameters,
$\rateSymbol_{\exLabel}\lesssim T_{2}^{-1}\lesssim\decohESReff$ is
satisfied.}
\end{figure}

The stationary photoluminescence $\Gamma=\photolum{-}+\photolum{+}$
consists of a $\sigma^{-}$ and a $\sigma^{+}$ polarized contribution,
$\photolum{-}=\Wem\rhoStat_{\Xd}$ and $\photolum{+}=\Wem\rhoStat_{\Xu}$,
respectively.  We find that the rates $\photolum{-}$ and $\photolum{+}$
are proportional to $\WeffESR/(\gamma+\WeffESR)$ for a given $\gamma$,
up to a constant background which is negligible for $\Wdu<\WeffESR$.
In particular, the total emission rate $\Gamma=\photolum{-}+\photolum{+}$
as function of $\detuningESReff$ is a Lorentzian with linewidth \begin{equation}
w=2\,\decohESReff\sqrt{1+\frac{\WmaxESR}{\gamma}},\end{equation}
 see also Fig.\  \ref{Fig: cwPL}. By analyzing the expression for
$\gamma$, we find the relevant parameter regime with the inequality
\begin{eqnarray}
w & \leq & 2\decohESReff\Bigg[1\!+\!\frac{2\WmaxESR}{\WeffEx}\left(1\!+\!\frac{\Wem}{\Wswitchoff}\!+\!\frac{\WXdXu}{\Wswitchoff}\right)\!\nonumber \\
 &  & +\!\frac{3\WmaxESR}{\Wswitchoff}\!+\!\frac{\WmaxESR}{\Wem}\left(1\!+\!\frac{3\,\WXdXu}{\Wswitchoff}\right)\Bigg]^{1/2}\!\!\!\!\!\!\!\!,\label{eqnLinewidth}\end{eqnarray}
which saturates for vanishing spin flip rates $\Wdu$ and $\Wud$.
Here, we have introduced the rate $\Wswitchoff=\WXuXd+\Wud\left(1+\Wem/\WeffEx\right)$
which describes different relaxation channels that lead to the ground
state $\spup$. These correspond to {}``switching off'' the laser
excitations because of Pauli blocking. The linewidth $w$ thus provides
a \emph{lower bound for} $T_{2}$: \begin{equation}
T_{2}\geq\decohESReff^{-1}\geq\frac{2}{w}.\end{equation}
Here, the second inequality saturates when the expression in brackets
in Eq. (\ref{eqnLinewidth}) becomes close to 1 (e.g., for efficient
hole spin relaxation \cite{Flissikowski,woods} $\Wswitchoff$ is
large and $w\approx2\,\decohESReff$). For the first inequality, $T_{2}\approx\decohESReff^{-1}$
for $\WeffEx^{\mathrm{max}}<1/T_{2}$, see Eq. (\ref{eqndecohESReff}).
To check our analytical approximation for $\Gamma$, we have solved
the generalized master equation numerically using the parameters given
in the caption of Fig.\  \ref{Fig: cwPL}. Comparing the two results
for $\Gamma$, we find that the relative difference is less than $0.2\:\%$.

Due to possible imperfections in the ODMR scheme described above,
e.g., due to mixing of hh and lh states or due to a small contribution
of the $\sigma^{+}$ polarization in the laser light, there can be
a small probability that the Pauli blocking of absorption is somewhat
lifted and the state $\spup$ can be optically excited. We describe
this process with the effective rate $\rateSymbol_{\exLabel,\uparrow}$.
It leads to an additional linewidth broadening, similar to the one
described with Eq.\ (\ref{eqnLinewidth}). We find that this effect
is small if $\rateSymbol_{\exLabel,\uparrow}<\WeffESR$.   

The setup discussed in this section combines optical excitation and
detection at the same wavelength. The laser stray light is an undesirable
background here and its detection can be avoided, e.g., by using a
polarization filter and by measuring only $\Gamma^{+}$. The laser
could also be distinguished from $\Gamma^{-}$ if  two-photon excitation
is applied, which is, e.g., possible with excitons in II-VI (e.g.,
CdSe \cite{cdse2phot} or CdS \cite{cds2phot}) and I-VII (e.g., CuCl
\cite{cucl2phot}) semiconductor nanocrystals. As another alternative,
the optical excitation could be tuned to an excited hole state (hh
or lh) \cite{grundmann}, possibly with a reversal of laser polarization.
Using a \emph{pulsed} laser would enable the distinction between luminescence
and laser light by time-gated detection.  See also Section \ref{sec:Spin-Rabi-Oscillations}
for another detection scheme with a pulsed laser. Another option is
to detect the resonant absorption instead of the photoluminescence,
using an optical transmission setup \cite{hoegele}. Finally, one
can also measure the photocurrent \cite{photocurrent,photocurrent2}
instead of the photoluminescence, which we discuss in the following
section.

\section{\label{sec:Read-Out-Via}Read Out via Photocurrent}

\begin{figure}[tb]
 \centerline{\includegraphics[%
  width=8.6cm,
  keepaspectratio]{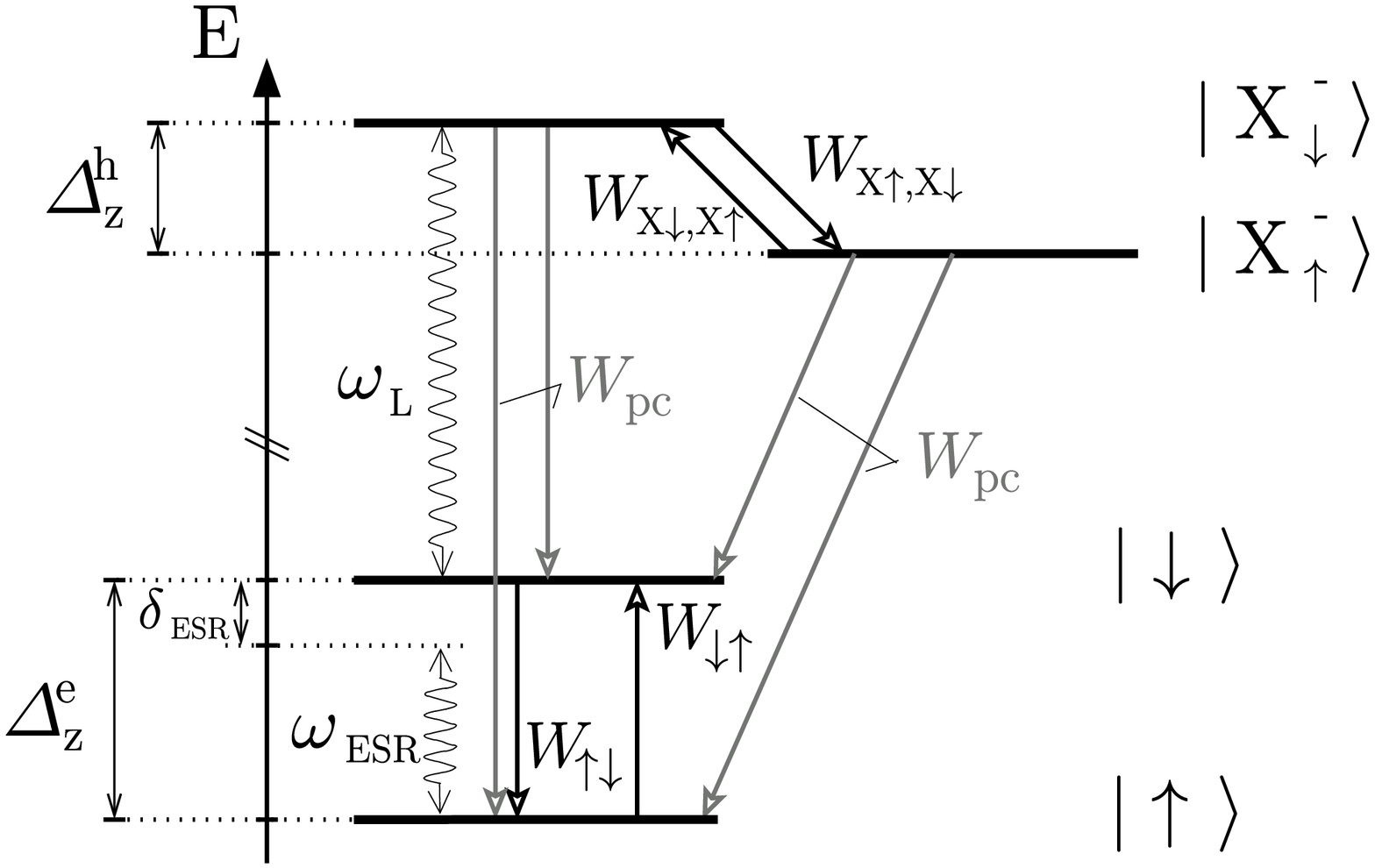} \vspace{1mm} }

\caption{Scheme of the transitions for the read out via photocurrent. The
tunneling of the electron and the hole out of the dot is spin-independent.
Therefore, transitions occuring at a rate $\Wpc$ lead from the charged
exciton states $\exMupket$, $\exMdownket$ to both spin states $\spup$
and $\spdown$, respectively. The remaining symbols are defined as
in Fig.\ \ref{fig:transitions}.\label{fig:pcstates}}
\end{figure}

As an alternative to photon detection, the presence of a charged exciton
$X^{-}$ on the dot can also be read out via an electric current (the
so-called photocurrent) \cite{photocurrent,photocurrent2}. Here,
a strong electric field is applied across the quantum dot, and one
electron and one hole tunnel out of the dot into two adjacent current
leads.  Thus, the total charge $e$ is transported through the dot
per optical excitation, where $e$ is the elementary charge. Because
the tunneling process is spin-independent, the remaining electron
on the dot has equal probabilities to be in state $\spup$ or in state
$\spdown$, in contrast to the read out using photoluminescence. 
We now calculate the stationary photocurrent. For this, we apply a
generalized master equation description, similarly as in Section \ref{sec:Generalized-Master-Equation}
for the photoluminescence. We introduce phenomenological photocurrent
rates $\Wpc$ as shown in Fig.\ \ref{fig:pcstates}. For strong tunneling
($\Wpc>\Wem$), optical recombination is negligible and the $X^{-}$
are predominantly detected via the photocurrent. The generalized master
equation is then given by%
\begin{eqnarray}
\rhoDotuu & = & \RabiESR{\textrm{Im}}\rhodu\!+\!\Wpc(\rho_{\XuXu}+\rho_{\XdXd})\!\nonumber \\
 &  & +\!\rateSymbol_{\ud}\rhodd\!-\!\rateSymbol_{\du}\rhouu,\label{eq:fullmastereqpc}\\
\rhoDotdd & = & -\RabiESR{\textrm{Im}}\rhodu+{\textrm{Im}}(\RabiEx^{*}\rho_{\Xdd})+\rateSymbol_{\du}\,\rhouu\nonumber \\
 &  & +\Wpc(\rho_{\XuXu}+\rho_{\XdXd})-\rateSymbol_{\ud}\,\rhodd,\\
\dot{\rho}_{\XdXd} & = & -{\textrm{Im}}(\RabiEx^{*}\rho_{\Xdd})+\rateSymbol_{\XdXu}\,\rho_{\XuXu}\nonumber \\
 &  & -\left(2\Wpc+\rateSymbol_{\XuXd}\right)\rho_{\XdXd},\\
\dot{\rho}_{\XuXu} & = & \rateSymbol_{\XuXd}\,\rho_{\XdXd}-\left(2\Wpc+\rateSymbol_{\XdXu}\right)\rho_{\XuXu},\\
\rhoDotdu & = & \frac{i}{2}\RabiESR\left(\rhodd-\rhouu\right)-\frac{i}{2}\RabiEx^{*}\rho_{\Xdu}\nonumber \\
 &  & -\left(i\detuningESR+T_{2}^{-1}\right)\,\rhodu,\\
\dot{\rho}_{\Xdu} & = & \frac{i}{2}\RabiESR\,\rho_{\Xdd}-\frac{i}{2}\RabiEx\rhodu\nonumber \\
 &  & -[i(\detuningESR+\detuningEx)+\decoherenceSymbol_{\Xdu}]\,\rho_{\Xdu},\\
\dot{\rho}_{\Xdd} & = & \frac{i}{2}\RabiESR\rho_{\Xdu}-\frac{i}{2}\RabiEx(\rhodd-\rho_{\XdXd})\nonumber \\
 &  & -(i\detuningEx+\decohEx)\rho_{\Xdd}.\end{eqnarray}
Note that in the previous expressions for $\decohEx$ and $\decoherenceSymbol_{\Xdu}$,
the relaxation rate $\Wem$ is now replaced by $\Wpceff=2\Wpc$. We
then obtain for the stationary solution \begin{eqnarray}
\rhoStat_{\uparrow} & = & \tilde{\eta}\WeffEx\Wpceff\!+\!\tilde{\eta}\WSud\,\Wpceff\!+\!\tilde{\eta}\WSud\!\!\left(\!\WeffEx\!+\!\Wpceff\!\right),\\
\rhoStat_{\downarrow} & = & \tilde{\eta}\WSdu\,\left(\WeffEx+\Wpceff\right)+\tilde{\eta}\WSdu\,\Wpceff,\\
\rhoStat_{\Xd} & = & \tilde{\eta}\WeffEx\,\WSdu,\\
\rhoStat_{\Xu} & = & \tilde{\eta}\WeffEx\,\WSdu.\end{eqnarray}
 Here, $\tilde{\eta}$ is a normalization factor such that $\sum_{n}\rho_{n}=1$.
The photocurrent $I_{\mathrm{pc}}=e\Wpceff(\rhoStat_{\Xd}+\rhoStat_{\Xu})$
is a Lorentzian as a function of the ESR detuning $\detuningESReff$.
The linewidth is bound by the inequality \begin{eqnarray}
w & \leq & 2\,\decohESReff\,\Bigg[1+4\WmaxESR\,\left(\frac{1}{\WeffEx}+\frac{1}{\Wpceff}\right)\Bigg]^{1/2},\end{eqnarray}
 where the right hand side is a smaller upper bound for $w$ than
the one obtained for the photoluminescence {[}Eq. (\ref{eqnLinewidth}){]}.
This can be understood by noting that above result for the photocurrent
can also be obtained from the expression for the stationary photoluminescence
(see Section \ref{sec:ESR-Linewidth-in}) by replacing $\Wem\rightarrow\Wpceff$
and in the limit $\rateSymbol_{\XuXd},\,\rateSymbol_{\XdXu}\rightarrow\infty$.

\section{Luminescence Intensity Autocorrelation Function\label{sec:Luminescence-Intensity-Autocorrelation}}

\begin{figure}[t]
\includegraphics[%
  width=8.6cm,
  keepaspectratio]{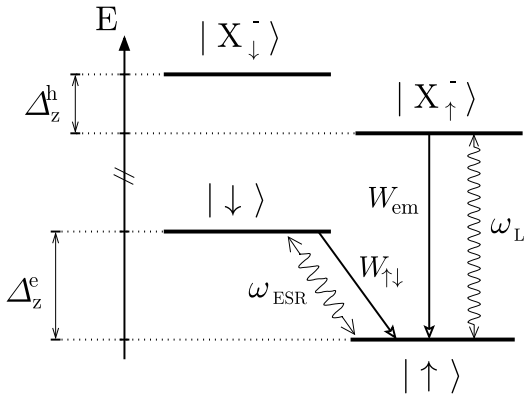}

\caption{\label{Fig: sigmapluslevels} Scheme of the transitions if an ESR
field and a $\sigma^{+}$ polarized laser field are applied. As described
in the text, $\exMdownket$ is decoupled from the other three states
at low temperatures. In this setup, the luminescence intensity autocorrelation
function $\left\langle I(t)I(t+\tau)\right\rangle $ can be used to
detect the decay of spin oscillations. Further, for a larger laser
intensity, $T_{1}$ can be measured as explained in Section \ref{sec:Luminescence-Intensity-Autocorrelation}.}
\end{figure}

The luminescence intensity autocorrelation function $\left\langle I(t)I(t+\tau)\right\rangle $
has recently been used in experiments to demonstrate the suitability
of single quantum dots for single-photon sources \cite{michler,zwiller}.
We discuss here that electron spin Rabi oscillations can be detected
via $\left\langle I(t)I(t+\tau)\right\rangle $. For this, we assume
that the laser polarization is changed to $\sigma^{+}$. At low temperatures
($kT<g_{hh}^{z}\mubohr B_{z}$, where $k$ is the Boltzmann constant),
excitations of the hole spin are negligible since $\WXdXu\ll\Wem$.
Then, the energetically highest state $\exMdownket$ is decoupled
from the three-level system $\spup$, $\spdown$, and $\exMupket$,
cf.\  Fig.\ \ref{Fig: sigmapluslevels}. After emission of a $\sigma^{+}$
photon, the dot is in the state $\spup$. For the transitions shown
in Fig.\ \ref{Fig: sigmapluslevels}, we derive a generalized master
equation similarly as Eqs.\ (\ref{eq:fullmastereq})-(\ref{eq:fullmastereqfin})
were derived according to Fig.\ \ref{fig:transitions}. We model
the time evolution of the dot state $\spup$ in lowest order in $\WeffEx$
and obtain the probability to be in the final state $\exMupket$ after
some time $\tau$. We consider the regime $\WeffEx\leq\decohESReff$
and $\decohESReff\ll\RabiESR<\Wem$ and obtain for the luminescence
intensity autocorrelation function\begin{equation}
\left\langle I(t)I(t+\tau)\right\rangle =\WeffEx^{2}\rhoStat_{\uparrow}^{\,2}P_{\uparrow}(\tau)+o(\WeffEx^{3}).\label{eq:LIA}\end{equation}
 Here, $I(t)$ is the $\sigma^{+}$ luminescence intensity, $\rhoStat_{\uparrow}\approx\WSud/(\WSud+\WSdu)$
is the stationary occupation of $\spup$, and $P_{\uparrow}(\tau)$
is the conditional probability to be again in the state $\spup$ after
the time $t+\tau$ if the state was $\spup$ at time $t$. For $\detuningESR=0$
and $\Wud=\Wdu$, we find  $P_{\uparrow}(\tau)\approx1/2+(1/2)\exp\left[-(\tau/2)\left(1/T_{2}+1/T_{1}\right)\right]\cos(\RabiESR\tau)$.
Thus, the inverse decay rate of the detected oscillations in $\left\langle I(t)I(t+\tau)\right\rangle $
{[}Eq.\ (\ref{eq:LIA}){]} gives a lower bound on $2T_{2}$. %

To conclude this section, we briefly point out that the single-spin
relaxation \emph{}time $T_{1}$ can be measured via a similar double
resonance scheme as discussed for $\left\langle I(t)I(t+\tau)\right\rangle $
above. This can be done in the regime $\RabiEx,\,\Wem\gg\RabiESR,\,\Wud$,
i.e., we require a larger intensity of the $\sigma^{+}$ laser as
considered for the $T_{2}$ measurement. Then, the system is predominantly
driven by the laser field. Occasionally, the ESR field excites the
electron spin and interrupts the optical excitations. After relaxation
of the spin, the laser again acts on the dot and gives rise to photoluminescence.
The mean time of photoluminescence interruptions due to ESR excitation
is thus given by $T_{1}$, similarly as for a single atom \cite{kimble}.

\section{\label{sec:Spin-Rabi-Oscillations}Spin Rabi Oscillations via Photoluminescence}

\begin{figure}[t]
\includegraphics[%
  width=8.6cm,
  keepaspectratio]{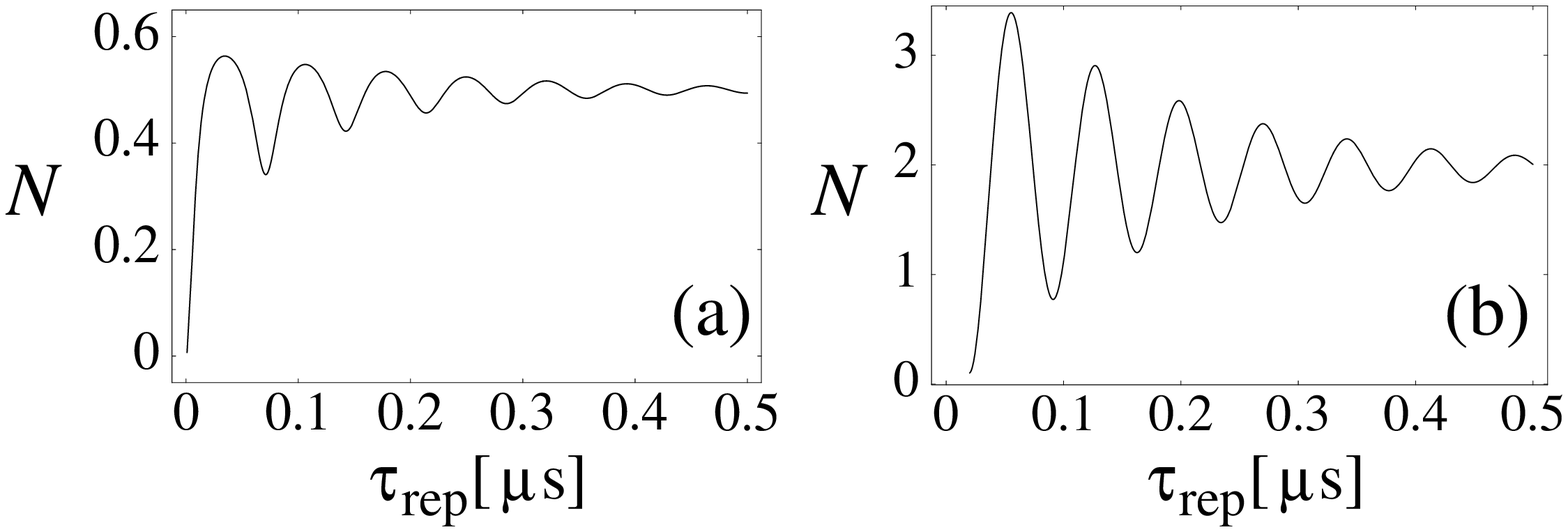}

\caption{\label{Fig: pulsedPL}The average number $N=\Gamma\reptime$ of photons
emitted per period $\reptime$ as a function of the laser pulse repetition
time $\reptime$. In (a), $\pi$ pulses are used for the laser with
$\Delta t=5\,\mathrm{ps}$ and $\RabiEx=\pi/\Delta t$. In (b), $N$
is shown for pulses with $\Delta t=20\,\mathrm{ns}$ and $\RabiEx=\pi/(500\,\mathrm{ps})$.
We have set $\detuningESR=0$. The other parameters are the same as
in Fig.\ \ref{Fig: cwPL}. The decay of the oscillation depends on
$T_{2}$. }
\end{figure}

The photoluminescence $\Gamma$  can be measured as a function of
the pulse repetition time $\reptime$ of a pulsed laser while keeping
$\detuningESR$ constant. We again consider cw ESR and  choose $\sigma^{-}$
for the laser polarization, while the previous restrictions on the
laser bandwidth still apply (see Section \ref{sec:Negatively-Charged-Excitons}).
 Since excessive population is trapped in the state $\spup$ during
a laser pulse due to hole spin flips and subsequent emission of a
photon, the dot is preferably in the state $\spup$ (rather than $\spdown$)
at the end of a laser pulse. During the {}``off'' time of the laser
between two pulses, the ESR field rotates the electron spin.  The
next laser pulse then reads out the spin state $\spdown$. Thus, as
a function of $\reptime$, the spin Rabi oscillations can be observed
in the photoluminescence (similarly as in $\left\langle I(t)I(t+\tau)\right\rangle $),
see Fig.\  \ref{Fig: pulsedPL}.  To model the pulsed laser excitation,
we consider square pulses of length $\Delta t$, for simplicity. In
the generalized master equation $\rhoDot(t)=\meq(t)\rho(t)$, we write
 $\meq(t)=\meqL$ {[}where $\meqL$ is defined via Eqs.\ (\ref{eq:fullmastereq})
- (\ref{eq:fullmastereqfin}){]} during a laser pulse and $\meq(t)=\meqnull$
otherwise, setting $\RabiEx=0$.  We obtain the steady-state density
matrix $\rhoinf$ of the dot state just after the pulse from $U_{p}\rhoinf=\rhoinf$,
where $U_{p}=\exp(\meqL\Delta t)\exp[\meqnull(\reptime-\Delta t)]$
describes the time evolution of $\rho$ during $\reptime$. %

The steady-state photoluminescence is now calculated by $\Gamma=\Wem\overline{(\rho_{\Xd}+\rho_{\Xu})}$,
where the bar symbolizes time averaging over many periods $\reptime$.
   If the laser pulse duration is longer than the lifetime of a
negatively charged exciton, $\Delta t>\rateSymbol_{\mathrm{em}}^{-1}$
(and not shorter than an optical $\pi$ pulse), the spin oscillations
become more pronounced, see Fig.\ \ref{Fig: pulsedPL} (b). This
is because after an optical recombination of the state $\exMdownket$,
the laser pulse is still on and excites the state $\spdown$ again
to $\exMdownket$. This iterated excitation increases the total probability
of a hole spin flip during a laser pulse and therefore the total population
trapped in the state $\spup$.

\section{\label{sec:Spin-precession-via}Spin Precession via Photoluminescence}

\begin{figure}[t]
\includegraphics[%
  width=7.8cm,
  keepaspectratio]{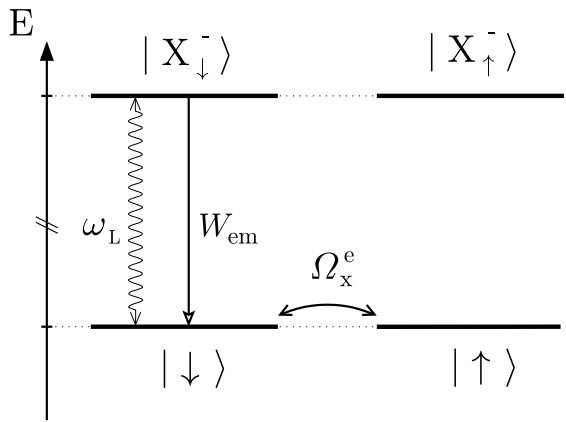}

\caption{\label{Fig: staticBlevels}Level scheme of the spin states in the
$z$ direction in the Voigt geometry. Optical transitions with circular
polarization occur vertically in this scheme. The transverse magnetic
field $B_{x}$ leads to spin precession, i.e., periodic oscillations
between the spin states $\spup$ and $\spdown$ at a frequency $\Omega_{x}^{e}=g_{e}^{x}\mubohr B_{x}/2$.
In this figure, we do not illustrate the precession of hole spins,
assuming that $\Wem\gg\Omega_{x}^{h}=g_{h}^{x}\mubohr B_{x}/2$.}
\end{figure}
Similar to Rabi oscillations, the precession of a single electron
spin in a static magnetic field can also be observed if pulsed laser
excitation is applied to a quantum dot charged with a single excess
electron. For this, we consider the Voigt geometry, i.e., a static
magnetic field is applied in a direction $x$, transverse to the laser
beam direction $z$. We again assume circular polarization of the
laser. Consequently, the optical transitions are between the spin
states along the quantization axis in $z$ direction, see Fig.\ \ref{Fig: staticBlevels}.
For low temperatures \cite{woods} and for $\Wem\gg\Omega_{x}^{h}=g_{h}^{x}\mubohr B_{x}/2$,
where $\Omega_{x}^{h}$ is the hole spin precession frequency, we
can neglect hole spin flips. Then, a state $\spdown$ is obtained
on the dot after the absorption of a $\sigma^{-}$ laser pulse and
subsequent optical recombination. This is not an eigenstate of the
quantum dot in the presence of the magnetic field $B_{x}$. In the
absence of an environment, the initial spin state $\spdown$ (at $t=0$)
evolves in time according to $\cos(\Omega_{x}^{e}t)\spdown-i\sin(\Omega_{x}^{e}t)\spup$
with precession frequency $\Omega_{x}^{e}=g_{e}^{x}\mubohr B_{x}/2$.
However, the spin precession decays due to decoherence. Using pulsed
laser excitation, the photoluminescence $\Gamma(\reptime)$ as function
of the pulse repetition time $\reptime$ oscillates according to the
spin precession and the damping is described by the spin decoherence
time, similarly as with ESR (see Sections \ref{sec:Luminescence-Intensity-Autocorrelation}
and \ref{sec:Spin-Rabi-Oscillations}). In the regime where the hole
spin flip rate is not small compared to $\Wem$, the visibility of
these photoluminescence oscillations is reduced, similarly as in Section
\ref{sec:Spin-Rabi-Oscillations}, where the spin polarization was
decreased for short laser pulses {[}see Fig.\ \ref{Fig: pulsedPL}
(a){]}. Finally, we note that in contrast to the detection of spin
Rabi oscillations (driven by ESR), in this setup spin decoherence
is measured in the absence of a driving field.

\section{\label{sec:Conclusions}Conclusions}

We have studied an ODMR setup with ESR and polarized optical excitation.
We have shown that this setup allows the optical measurement of the
single electron spin decoherence time $T_{2}$ in semiconductor quantum
dots. The discussed cw and pulsed optical detection schemes can also
be combined with pulsed instead of cw ESR, allowing spin echo and
similar standard techniques.  Such pulses can, e.g., be produced
via the AC Stark effect \cite{ACstark,pryor}. Further, as an alternative
to photoluminescence detection, photocurrent can be used to read out
the charged exciton, and the same ODMR scheme can be applied. We have
finally described a scheme, where single spin precession can be detected
via photoluminescence in a similar excitation setup.

\section*{Acknowledgments}

We thank M.H. Baier, J.C. Egues, A. Högele, A.V. Khaetskii, B. Hecht,
and H. Schaefers for discussions. We acknowledge support from the
Swiss NSF, NCCR Nanoscience, EU Spintronics, DARPA, ARO, and ONR.

\end{document}